\documentclass[english,twocolumn,showpacs,preprintnumbers]{revtex4}
\usepackage{times}
\usepackage[T1]{fontenc}
\usepackage[latin1]{inputenc}
\usepackage{prettyref}
\usepackage{amsmath}
\usepackage{graphicx}
\usepackage{amssymb}

\makeatletter
\usepackage{babel}
\makeatother
\begin{document}

\title{Shock wave theory for rupture of rubber}

\author{M. Marder}

\affiliation{Center for Nonlinear Dynamics and Department of Physics, The University
of Texas at Austin, Austin TX 78712}

\date{June 20, 2004}


\begin{abstract}
This article presents a theory for the rupture of rubber. Unlike conventional
cracks, ruptures in rubber travel faster than the speed of sound,
and consist in two oblique shocks that meet at a point. Physical features
of rubber needed for this phenomenon include Kelvin dissipation and
an increase of toughness as rubber retracts. There are three levels
of theoretical description: an approximate continuum theory, an exact
analytical solution of a slightly simplified discrete problem, and
numerical solution of realistic and fully nonlinear equations of motion.
\end{abstract}
\maketitle

\paragraph*{Introduction-- }

Rapidly moving cracks in brittle materials under tension have a number
of common characteristics. They cannot move faster than the shear
wave speed\cite{Freund.90,Broberg.99,Ravi.03}, and often exhibit
a limiting velocity around half that value because of instabilities
of the crack tip\cite{Fineberg.99}. Stresses near the tip rise in
a universal singularity as $1/\sqrt{r}$. In this Letter, I show that
ruptures in rubber are different. They are supersonic. There is stress
enhancement but no stress singularity near their tips. They constitute
a new sort of failure mode that combines characteristics of shocks
and cracks.

The motivation for this study comes from experiments showing that
cracks in rubber travel faster than the shear wave speed, and that
the tip has a wedge--like shape resembling a shock\cite{Petersan.04}.
Planar shock fronts in rubber were previously observed by Kolsky\cite{Kolsky.69}.
It has not been clear how to interpret the experiments because the
large nonlinearities of rubber invalidate immediate comparison with
the customary theory of linear elastic fracture mechanics. In particular,
one assumption of conventional fracture mechanics is that material
ahead of a crack tip is strained by a vanishingly small amount, while
in popping rubber the strains are several hundred percent.

Intersonic tensile cracks have been observed in numerics of Buehler,
Abraham and Gao\cite{Buehler.03}. In their calculations, this behavior
is produced by a rise in sound speed near the crack tip. Here the
mechanism is different; there is no rise in sound speed. Instead,
two other physical ingredients work together both in numerical simulations
and in analytical calculations to reproduce the basic experimental
observations. First and most important, the equation of motion for
rubber includes dissipation of the Kelvin form; Langer\cite{Langer.92}
has observed that such terms may permit supersonic motion. Second,
the rubber must be able to sustain larger stresses when it is relaxed
along one axis than when it is stretched equally in all directions.

\paragraph*{Continuum Theory of Rubber--}

Strains in rubber are several hundred percent at rupture and one must
use nonlinear elastic theory to describe the situation. Sound speeds
in rubber are adequately described\cite{Petersan.04} by one of the
most familiar free energies for non--linear elastic solids, the one
due to Mooney and Rivlin\cite{Treloar.75,Mooney.40,Rivlin.48}. For
this free energy, define the Lagrangean strain tensor\cite{Eringen.74}

\begin{equation}
E_{\alpha\beta}\equiv\mbox{$\frac{1}{2}$}\left[\sum_{\gamma}\frac{\partial u_{\gamma}}{\partial r_{\alpha}}\frac{\partial u_{\gamma}}{\partial r_{\beta}}-\delta_{\alpha\beta}\right].\label{eq:Eab}\end{equation}
Here $\vec{u}(\vec{r})$ describes the distance from the origin of
a mass point that was located at $\vec{r}$ before the rubber was
stretched up. From this strain tensor one can define three rotationally
invariant quantities, which are $I_{1}^{3D}={\rm Tr\,}E$, $I_{2}^{3D}=\sum_{\alpha<\beta}\left[E_{\alpha\alpha}E_{\beta\beta}-E_{\alpha\beta}^{2}\right]$,
and $I_{3}^{3D}={\rm det}\, E$. The Mooney--Rivlin theory says that
the free energy density of rubber is\begin{equation}
U/\rho\equiv w=a(I_{1}^{3D}+bI_{2}^{3D}),\end{equation}
where $U$ has units of energy per volume, $\rho$ is mass density,
$a$ is a constant with units of velocity squared, and $b$ is dimensionless.
For a thin sheet of rubber, one can replace the three--dimensional
theory by an effective two--dimensional one, using the facts that
rubber is highly incompressible\cite{Treloar.75}, and that one can
neglect all the components of the strain tensor $E_{\alpha z}$ except
for $E_{zz}$. In two dimensions one has only two invariants,\begin{equation}
I_{1}=E_{xx}+E_{yy};\quad I_{2}=E_{xx}E_{yy}-E_{xy}^{2},\label{eq:I1I2}\end{equation}
and using incompressibility to solve for $E_{zz}$ one finds\begin{equation}
E_{zz}=\mbox{$\frac{1}{2}$}\left(\frac{1}{4I_{2}+2I_{1}+1}-1\right).\label{eq:Ezz}\end{equation}
Thus one obtains an effective two--dimensional Mooney--Rivlin theory\begin{equation}
w(I_{1},I_{2})=a\left(I_{1}+bI_{2}+E_{zz}(1+bI_{2})\right).\label{eq:w}\end{equation}
For large strains, $E_{zz}$ becomes negligibly small compared to
$E_{xx}$ or $E_{yy}$. However, as rubber relaxes to equilibrium,
the terms proportional to $E_{zz}$ become important. They are what
ensure that $\vec{u}=\vec{r}$ is a minimum energy state.

For studying the rupture of rubber, the energy density in Eq. \prettyref{eq:w}
is both too simple and too complicated. It is too simple because it
does not account for the fact that when rubber is stretched enough,
the polymers pull apart and the force between adjacent regions drops
irreversibly to zero. It is too complicated because the terms involving
$I_{2}$ and $E_{zz}$ produce nonlinear equations of motion that
are impossible to solve analytically. Therefore, to analyze the problem,
I will pursue two different routes. First, I will discuss numerical
routines that supplement Eq. \prettyref{eq:w} with information about
rupture, toughening, and dissipation, and produce supersonic solutions.
Second, I will isolate from Eq. \prettyref{eq:w} terms that are sufficient
to produce good agreement with numerics and experiment, while simplifying
matters enough to permit analytical solution.

\paragraph*{Numerical System--}

To study rubber rupture numerically, consider a collection of mass
points $u_{i}$ whose equilibrium locations lie on a triangular lattice,
and that are connected with bonds to nearest neighbors. Take the lattice
spacing of the unstretched configuration to be $\Delta.$ For numerical
representation of the strain invariants, let $\vec{u_{ij}}\equiv\vec{u_{j}}-\vec{u}_{i},$
let $n(i)$ refer to the nearest neighbors of $i$, and define\begin{subequations}

\begin{equation}
F_{i}=\frac{1}{6}\sum_{j\in n(i)}\begin{cases}
\left(\vec{u}_{ij}\cdot\vec{u}_{ij}-\Delta^{2}\right) & \textrm{if }u_{ij}<\lambda_{f}\\
\lambda_{f}^{2}-\Delta^{2} & \textrm{else}\end{cases}\label{eq:MO1}\end{equation}

\begin{equation}
G_{i}=\frac{1}{9}\sum_{j\in n(i)}\begin{cases}
\left(\vec{u}_{ij}\cdot\vec{u}_{ij}-\Delta^{2}\right)^{2} & \textrm{if }u_{ij}<\lambda_{f}\\
\left(\lambda_{f}^{2}-\Delta^{2}\right)^{2} & \textrm{else}\end{cases}\label{eq:MO2}\end{equation}
\begin{equation}
H_{i}=\frac{1}{27}\sum_{j\neq k\in n(i)}h(u_{ij})h(u_{ik})\left(\vec{u}_{ij}\cdot\vec{u}_{ik}+2\Delta^{2}\right)^{2},\label{eq:MO3}\end{equation}
\begin{equation}
\mbox{and}\quad h(u)=1/(1+e^{(u-u_{c})/u_{s}}).\label{eq:MO4}\end{equation}

\label{eq:MO}\end{subequations}From these numerical quantities, one
can form representations of the strain invariants as follows:

\begin{subequations}

\begin{equation}
I_{1}^{i}=F_{i}/\Delta^{2}\label{eq:J1i}\end{equation}

\begin{equation}
I_{2}^{i}=(9/8)\left(G_{i}-H_{i}+4\right)/\Delta^{4}.\label{eq:J2ib}\end{equation}

\label{eq:I1I2i}\end{subequations}and finally construct the energy
from\begin{equation}
U=\sum_{i}mw(I_{1}^{i},I_{2}^{i}),\label{eq:energy}\end{equation}
where $m$ is the mass in a unit cell, and the energy density $w$
is given by Eq. \prettyref{eq:w}. The quantities in \prettyref{eq:MO}
are chosen according to two ideas. First, they are designed so that
when all bonds at a node are shorter than a critical failure extension
$\lambda_{f}$, in the continuum approximation Eqs. \prettyref{eq:I1I2i}
reproduce the strain invariants in Eq. \prettyref{eq:I1I2}. Second,
they are designed so that when bonds are stretched to an extension
greater than $\lambda_{f},$ they break. For the three--body term
in Eq. \prettyref{eq:MO3}, it is necessary to introduce a soft cutoff
through the function $h$ described in Eq. \prettyref{eq:MO4}, in
which $u_{s}$ is a parameter on the order of 0.1 that sets the scale
over which contributions to the three--body term drop to zero.

\begin{figure}
\includegraphics[%
  width=1.0\columnwidth]{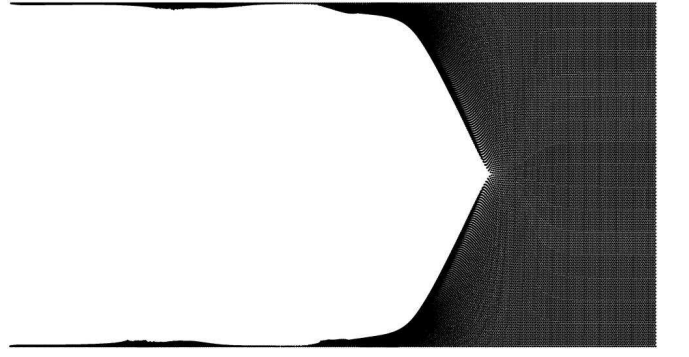}

\caption{Solution of Eq. \prettyref{eq:motion}, with experimental values
of $a=501$ $\textrm{m}^{2}/\textrm{s}^{2}$, $b=.1$06, Kelvin dissipation
$\beta=3$, and rupture extension $\lambda_{f}$= 5.5. Before rupture,
the system is stretched vertically by a factor of $\lambda_{y}=3.2$,
and horizontally by a factor of $\lambda_{x}=2$.1. The system has
been allowed to run for 500 time units, by which time it is approaching
a steady state, apart from bunching up of material as at collides
with rigid supports at top and bottom of the system. The rupture is
able to run as long as needed to the right by pasting new material
on the right and discarding it from the left. Note the wedge-like
shape of the rupture. The (Lagrangean) speed of shear waves ahead
of the rupture is $c=21.8$ m/s, and the rupture travels at a speed
$v=24.9$ m/s. The system is 200 rows high with 70,000 particles,
and in its unstretched configuration is twice as wide as it is tall.\label{cap:Solution-of-Eq.}}
\end{figure}

Figure \ref{cap:Solution-of-Eq.} shows an image of a steady state
obtained by solving dynamical equations that follow from Eq. \prettyref{eq:energy}.
The precise equation of motion includes dissipation of the Kelvin
form, and is \begin{equation}
m\ddot{u}_{i}^{\alpha}=-\partial U/\partial u_{i}^{\alpha}+\sum_{j\in n(i)}\frac{a\beta}{3}\dot{u}_{ij}^{\alpha}\theta(\lambda_{f}-u_{ij}).\label{eq:motion}\end{equation}
One final rule is employed. Whenever some bond $u_{ij}$ drops to
a length less than $1.5\Delta$, the failure extension $\lambda_{f}$
for the remaining bonds attached to nodes $i$ and $j$ increases.
Without some rule of this type, the back faces of the crack disintegrate.
Essentially, the back faces of the rupture act like a string under
tension pulling bonds at the tip apart, and they must be able to sustain
tensions sufficient to do so; for details, see Eq. \ref{eq:back_rupture}.

Numerical solutions of Eq. \prettyref{eq:motion} agree acceptably
with experiment. I have tried to determine which terms in it are really
needed. Progressively stripping elements from \prettyref{eq:motion}
I found what is most likely the simplest set of equations supporting
supersonic solutions. These explain the nature of the solutions, and
the conditions under which they arise.

\paragraph*{Neo--Hookean Continuum Theory--}

Experimentally\cite{Petersan.04}, the dimensionless parameter $b$
in Eq. \prettyref{eq:w} is .106, so in a first theoretical account
one can set $b=0$. For strains large enough also to neglect $E_{zz}$,
Eq. \prettyref{eq:w} reduces (up to an additive constant) to the
Neo--Hookean energy density\begin{align}
w & =aI_{1}=\frac{\rho c^{2}}{2}\left[\left({\textstyle \frac{\partial u_{x}}{\partial x}}\right)^{2}+\left({\textstyle \frac{\partial u_{y}}{\partial x}}\right)^{2}+\left({\textstyle \frac{\partial u_{x}}{\partial y}}\right)^{2}+\left({\textstyle \frac{\partial u_{y}}{\partial y}}\right)^{2}\right]\label{eq:NeoHookean}\end{align}
The equation of motion that follows from this energy is (for $\alpha=x$
or $y$ )\begin{equation}
\ddot{u}_{\alpha}=c^{2}\nabla^{2}u_{\alpha}.\label{eq:wave}\end{equation}
Despite the fact that the equation of motion Eq. \prettyref{eq:wave}
is an ordinary wave equation, it describes large extensions. The most
undesirable feature of this theory is that its ground state consists
in material that has collapsed down to a point; this results from
dropping the terms proportional to $E_{zz}$ from Eq. \prettyref{eq:w}.
However, as shown in Figure \ref{cap:Comparison-of-theory,} rupture
speeds are essentially unaffected by the presence of these terms.
The theory has the great advantage that it can be solved exactly.
For crack--like solutions, with Lagrangean variables $(x,y)=\vec{r}$
one has for $y=0$ and $x<0$ the boundary condition $\partial u_{y}/\partial y=0.$
Neither this boundary condition nor the equations of motion couple
$u_{x}$ and $u_{y}$; therefore, the equations support solutions
where $u_{x}=\lambda_{x}x$ does not change in time, and the motion
of the mass points is purely vertical. These solutions are identical
to the solutions for a crack in anti-plane shear\cite{Klingbeil.66},
as recorded for example in Ref. \cite{Freund.90}, p. 356. The static
solution has a parabolic tip. Steady states moving at velocity $v$,
are identical to the static solution, but are Lorenz contracted in
the direction of motion by a factor of $\sqrt{{1-v^{2}/c^{2}}}$.
As the crack speed $v$ approaches the wave speed $c$, the tip becomes
increasingly blunt. 

The wave speed $c$ has been thought the upper speed limit for crack--like
solutions of Eq. \prettyref{eq:wave}. However, supersonic solutions
are possible if one adds Kelvin dissipation corresponding to the rightmost
term in Eq. \prettyref{eq:motion} to obtain for a steady state moving
at velocity $v$,

\begin{subequations}\begin{equation}
v^{2}\frac{\partial^{2}u}{\partial x^{2}}=c^{2}\nabla^{2}u-vc^{2}\beta\nabla^{2}\frac{\partial u}{\partial x}.\label{eq:motion2a}\end{equation}
The variable $u$ in this equation is the vertical motion of mass
points $u_{y}$; the horizontal locations of all mass points remain
fixed at $u_{x}=\lambda_{x}x,$ so there is no need to keep track
of them further. Supplement Eq. \prettyref{eq:motion2a} with the
boundary conditions\begin{equation}
\frac{\partial u}{\partial y}=v\beta\frac{\partial^{2}u}{\partial x\partial y}\quad\textrm{for}\quad\  x<0;\quad u=0\quad\textrm{for}\quad x>0.\label{eq:motion2b}\end{equation}
\begin{equation}
u\rightarrow\lambda_{y}y\quad\textrm{as}\quad y\rightarrow\infty.\label{eq:motion2c}\end{equation}
\label{eq:motion2}\end{subequations}Solutions of Eq. \ref{eq:motion2}
can be obtained with the Wiener--Hopf technique\cite{Noble.58}. One
has the following results for the upper face of the rupture where
$y=0^{+}$ and $x$<0: \begin{equation}
\frac{\partial u}{\partial x}\Big\vert_{y=0}=-\int_{x}^{0}dx'\:\frac{\lambda_{y}e^{x/v\beta}}{\sqrt{-\pi v\beta x(v^{2}/c^{2}-1)}},\end{equation}
and

\begin{equation}
\left.\frac{\partial u}{\partial y}\right|_{y=0}=\frac{\lambda_{y}e^{x/v\beta}v/c}{\sqrt{v^{2}/c^{2}-1}}.\label{eq:strain}\end{equation}
Therefore, the slope $\alpha$ of the back face of the rupture seen
in the lab is\begin{subequations}\begin{equation}
\frac{-\lambda_{y}}{\lambda_{x}\sqrt{{v^{2}/c^{2}-1}}}.\label{eq:slope}\end{equation}
This is the slope of a shock cone trailing an object traveling at
speed $v>c$ in a medium of wave speed $c.$ Note that the velocities
$v$ and $c$ are measured in a Lagrangean reference frame described
by variables $x$ and $y.$ Horizontal and vertical speeds measured
in the laboratory are larger by factors of $\lambda_{x}$ and $\lambda_{y}$
respectively; this geometrical fact accounts for the factor $\lambda_{y}/\lambda_{x}$
in \prettyref{eq:slope}. The vertical strain at the origin is obtained
by setting $x=0$ in Eq. \prettyref{eq:strain}. One obtains a simple
but approximate prediction for rupture speed by checking when bonds
angled at 60$^{\circ}$ in a triangular lattice reach their breaking
point $\lambda_{f}:$\begin{eqnarray}
\lambda_{f}^{2} & = & \frac{1}{4}\lambda_{x}^{2}+\frac{3}{4}\left[\frac{\partial u}{\partial x}\Big\vert_{(0,0)}\right]^{2}\label{eq:failure}\\
 & \Rightarrow & \frac{\lambda_{y}}{\sqrt{(4\lambda_{f}^{2}-\lambda_{x}^{2})/3}}=\sqrt{1-c^{2}/v^{2}}\end{eqnarray}

\label{eq:predictions}\end{subequations} In order to compare with
experiment, there is a single free parameter to fix, which is the
breaking point $\lambda_{f}.$ Figure \ref{cap:Comparison-of-theory,}
shows a comparison of the predictions from Eqs. \ref{eq:predictions}
with experimental and numerical data, using $\lambda_{f}=5.5.$

\begin{figure}[t]
\includegraphics[%
  width=1.0\columnwidth,
  keepaspectratio]{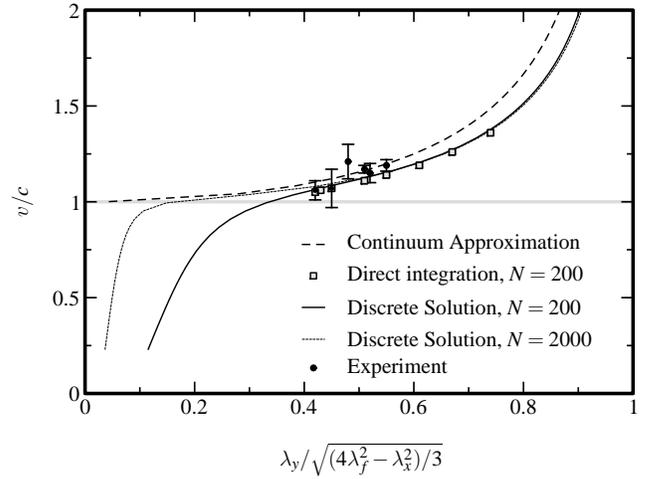}

\caption{Comparison of theory, experiment, and numerics for rubber rupture
velocities. Experimental velocities are scaled by $c\lambda_{x}$,
with $c=$22 m/s, while the vertical extension $\lambda_{y}$ is scaled
by $\sqrt{(4\lambda_{f}^{2}-\lambda_{x}^{2})/3}$. The continuum approximation
is given in Eq. \prettyref{eq:failure}. Direct integration of Eq.
\prettyref{eq:motion} is carried out in triangular lattices $N=$200
rows high in the Neo--Hookean limit where $b=0,$ with Kelvin dissipation
$\beta=3,$ and retaining $E_{zz}$ as in Eq. \prettyref{eq:w}. The
Discrete Analytical Solution is an exact solution of the same system
using the Wiener-Hopf technique,with the three differences. First,
$E_{zz}$ is neglected in the analytical solution. Second, the analytical
system is infinitely long in the horizontal direction, while the numerical
system is finite. Third, in the numerical system there is a brief
time when only one of two crack-line bonds has snapped, and horizontal
forces on crack-line atoms do not balance to zero, while in the analytical
solutions, all forces in the horizontal direction are ignored. Analytical
solutions for systems both 200 and 2000 rows high are displayed to
show how the continuum limit is achieved. Experimental results courtesy
of Paul Petersan and Robert Deegan. \label{cap:Comparison-of-theory,}}
\end{figure}

An additional interesting quantity to check is the distance squared
between horizontal mass points behind the rupture. It is\begin{equation}
\lambda_{x}^{2}+\frac{\lambda_{y}^{2}}{v^{2}/c^{2}-1}=\frac{4}{3}\lambda_{f}^{2}+\frac{2}{3}\lambda_{x}^{2}-\lambda_{y}^{2}.\label{eq:back_rupture}\end{equation}
 This quantity exceeds $\lambda_{f}^{2}$ for characteristic values
of $\lambda_{f},$ $\lambda_{x},$ and $\lambda_{y},$ which explains
why it is necessary for $\lambda_{f}$ to increase behind the rupture
if the back surface is not to disintegrate.

\paragraph*{Neo--Hookean Discrete Theory--}

Not only can the continuum Neo--Hookean theory be solved, but the
discrete theory, Eq. \prettyref{eq:motion} can also be solved exactly,
provided in Eq. \prettyref{eq:w} one sets $b=0$ and $E_{zz}=0.$
The solution involves the application of methods described in Refs.
\cite{Slepyan.02,Heizler.02,Marder.95.jmps}, and details will be
presented elsewhere. Figure \ref{cap:Comparison-of-theory,} shows
exact solutions for rupture speeds in systems 200 and 2000 rows high
compared both with direct integration of the equations of motion and
experiment. In addition to removing discrepancies between the very
simple results in Eqs. \prettyref{eq:predictions} and numerics, solving
the discrete model explains the conditions under which one gets supersonic
or subsonic solutions for cracks in tension. 

The basic result is this: including dissipation through $\beta$ in
the equation of motion introduces a length scale $\beta c$ into the
problem. The behavior of cracks hinges on the ratio of $\beta c$
to the lattice spacing $\Delta$. When $\beta c/\Delta$ is much less
than one, cracks behave as in conventional fracture mechanics, and
their speed is limited from above by $c,$ except within a very narrow
window of strains where all bonds in the system ahead of the crack
approach their breaking point. As $\beta c/\Delta$ approaches and
exceeds one, dissipation progressively destroys the stress singularity
around conventional crack solutions, but at the same time it permits
the appearance of supersonic solutions. Note in Eq. \prettyref{eq:failure}
that rupture speed is determined by vertical extension $\lambda_{y},$
rather than by the total energy stored ahead of the crack tip as in
conventional fracture mechanics. Exact solution of the discrete Neo--Hookean
theory shows that \prettyref{eq:failure} is not completely accurate,
but its scaling properties are correct. One sees in Figure \ref{cap:Comparison-of-theory,}
that the relation between rupture velocity and system extension $\lambda_{y}$
has essentially reached the macroscopic limit for systems 200 rows
high and velocities $v$ above $1.05c$. The macroscopic limit is
subtle near $v=c,$ since solutions with speeds above and below $c$
scale differently as system size goes to infinity. 

Establishing the existence of supersonic ruptures in tension opens
up many possibilities for future work. The supersonic ruptures in
experiment begin to oscillate once $\lambda_{x}$ exceeds a critical
value. The numerical and analytical tools provided here should provide
an appropriate starting point for studying the oscillations. Finally,
it would be interesting to know if there are materials different from
rubber that meet the conditions needed to sustain supersonic ruptures.

\begin{acknowledgments}
I am indebted to Jim Rice for pointing out, in a lengthy email, that
it would be profitable to study this problem with the Neo--Hookean
theory. I have had many discussions about the physics with Robert
Deegan, Paul Petersan, and Harry Swinney. Financial support from the
National Science Foundation through DMR-0401766 is gratefully acknowledged.
\end{acknowledgments}
\bibliographystyle{/usr/share/texmf/bibtex/bst/revtex4/apsrev}
\bibliography{/home/marder/crack/tex/fracture,/home/marder/crack/tex/rubber/local}

\end{document}